\documentclass[reprint, aps, prl, superscriptaddress,nobibnotes]{revtex4-1}

\usepackage{graphicx}
\usepackage{dcolumn}
\usepackage{amsmath}
\usepackage{textcomp}
\usepackage{bm}
\usepackage{braket}
\usepackage[left]{lineno}
\usepackage{color,soul}

\begin{document}

\title{Effective g factor of low-density two-dimensional holes in a Ge quantum well}

\author{T. M. Lu}
\email{tlu@sandia.gov}
\affiliation{Sandia National Laboratories, Albuquerque, New Mexico 87185, USA}
\author{C. T. Harris}
\affiliation{Sandia National Laboratories, Albuquerque, New Mexico 87185, USA}
\affiliation{Center for Integrated Nanotechnologies, Sandia National Laboratories, Albuquerque, New Mexico, 87123, USA}
\author{S.-H. Huang}
\affiliation{Department of Electrical Engineering and Graduate Institute of Electronic Engineering, National Taiwan University, Taipei 10617, Taiwan, R.O.C.}
\affiliation{National Nano Device Laboratories, Hsinchu 30077, Taiwan, R.O.C.}
\author{Y. Chuang}
\affiliation{Department of Electrical Engineering and Graduate Institute of Electronic Engineering, National Taiwan University, Taipei 10617, Taiwan, R.O.C.}
\affiliation{National Nano Device Laboratories, Hsinchu 30077, Taiwan, R.O.C.}
\author{J.-Y. Li}
\affiliation{Department of Electrical Engineering and Graduate Institute of Electronic Engineering, National Taiwan University, Taipei 10617, Taiwan, R.O.C.}
\affiliation{National Nano Device Laboratories, Hsinchu 30077, Taiwan, R.O.C.}
\author{C. W. Liu}
\affiliation{Department of Electrical Engineering and Graduate Institute of Electronic Engineering, National Taiwan University, Taipei 10617, Taiwan, R.O.C.}
\affiliation{National Nano Device Laboratories, Hsinchu 30077, Taiwan, R.O.C.}

\date{\today}

\begin{abstract}

We report measurements of the effective $g$ factor of low-density two-dimensional holes in a Ge quantum well.  
Using the temperature dependence of the Shubnikov-de Haas oscillations, we extract the effective $g$ factor in a magnetic field perpendicular to the sample surface. 
Very large values of the effective $g$ factor, ranging from $\sim13$ to $\sim28$, are observed in the density range of $1.4\times10^{10}$ cm$^{-2}$ to $1.4\times10^{11}$ cm$^{-2}$.
When the magnetic field is oriented parallel to the sample surface, the effective $g$ factor is obtained from a protrusion in the magneto-resistance data that signifies full spin polarization.
In the latter orientation, a small effective $g$ factor, $\sim1.3-1.4$, is measured in the density range of $1.5\times10^{10}$ cm$^{-2}$ to $2\times10^{10}$ cm$^{-2}$.
This very strong anisotropy is consistent with theoretical predictions and previous measurements in other 2D hole systems, such as InGaAs and GaSb.

\end{abstract}

\maketitle

In recent years, Ge has emerged as a versatile electronic material for both industrial applications and fundamental physical studies \cite{Pillarisetty2011,Shi2015a,Shi2015b,Mironov2016, Lu2017}.
Possessing a very high hole mobility at room temperature and being epitaxially compatible with Si, Ge is considered a promising channel material for future digital electronics \cite{Zhang2013}.
A cubic Rashba spin-orbit coupling was recently reported for two-dimensional (2D) holes in a Ge quantum well, making Ge a candidate material for spintronics as well \cite{Moriya2014}.
With recent advances in Ge/SiGe epitaxy, the low-temperature mobility of two-dimensional (2D) holes can now constantly reach $1\times 10^{5}$ cm$^{2}$V$^{-1}$s$^{-1}$ \cite{Dobbie2012,Laroche2016}, enabling refined research in the quantum regime, such as the fractional quantum Hall effect \cite{Shi2015a,Mironov2016} and density-controlled quantum Hall ferromagnetic transition \cite{Lu2017}, which were previously more difficult or impossible to study due to disorder in the material.

Two important material parameters of a 2D system are the effective mass, $m^*$, and the effective $g$ factor, $g^*$.  
These parameters determine how a 2D system in a specific material responds to external electric and magnetic fields, and, in turn, they indicate device performance quality and foretell the likelihood of interesting physical phenomena arising. 
Both $m^*$ and $g^*$ derive from the bulk band structure and quantum confinement, with their values being heavily influenced by carrier interactions.
For Ge 2D holes, most measurements yield $m^*\sim0.07-0.1$ $m_0$, where $m_0$ is the free electron mass \cite{Irisawa2003,Rossner2003,Dobbie2012,Hassan2013,Failla2015,Failla2016}.
However, $g^*$ for Ge 2D holes has not been studied extensively.
In general, theoretical considerations conclude that $g^*$ is highly anisotropic, with the effective in-plane $g$ factor for magnetic fields parallel to the 2D plane, $g_{ip}$, being vanishingly small and the effective perpendicular $g$ factor, $g_{p}$, being on the order of $20$  based on bulk Luttinger parameters for Ge \cite{Luttinger1956,Winkler2000,Nenashev2003,Winkler2003}.
Compared to theoretical calculations, the values of $g_p$ measured by optical techniques are typically much smaller and range from $2.8$ to $7.0$ \cite{Failla2015,Failla2016}.
However, to date, very little experimental work to quantify effective $g$ factor anisotropy exists.
Lastly, the density ($p$) dependencies of $g_p$ and $g_{ip}$ have not been systematically studied, especially for low-density values when $p$ is in the $10^{10}$ cm$^{-2}$ regime, where carrier interactions are important.
Using an undoped Ge/SiGe heterostructure field-effect transistor (HFET) architecture, we report in this Letter our measurements of $g_p$ and $g_{ip}$ in the low-density regime. 
A very strong anisotropy is observed, with $g_{ip}\sim1.3-1.4$ and $g_{p}$ at least an order of magnitude higher.
Carrier interactions enhance $g_p$ as $p$ is decreased, and $g_p$ reaches $\sim30$ at $p=1.4\times10^{10}$ cm$^{-2}$.

The HFET device used in this study was fabricated using a $484$-nm-deep Ge/Si$_{0.2}$Ge$_{0.8}$ quantum well heterostructure.
This starting material was identical to that used in our previous work \cite{Laroche2016}, and its material characterizations and device transport characteristics have been reported therein.
The most prominent feature of an HFET comprised of this material is the remarkably low densities it possesses, in the $10^{10}$ cm$^{-2}$ range, which is accessible through electrical transport measurements.
Using this device, we performed standard, low-frequency lock-in measurements to obtain the longitudinal magneto-resistance ($R_{xx}$) and the transverse magneto-resistance ($R_{xy}$) in perpendicular magnetic fields in a pumped helium cryostat with a base temperature $T=1$ K, and in-plane magneto-resistance ($R_{ip}$) in parallel magnetic fields in a pumped $^3$He cryostat with a base temperature $T=0.3$ K.
The temperature dependence of the Shubnikov-de Haas (SdH) oscillations in $R_{xx}$ allows us to extract the Zeeman gaps, from which $g_p$ can be obtained.
For in-plane magnetic fields, the Zeeman effect manifests as a protrusion in the $R_{ip}$.  The Zeeman splitting and, in turn, $g_{ip}$, can be extracted from the magnetic field where this resistance jut occurs.

\begin{figure}[h]
\resizebox{3.3 in}{!}{\includegraphics{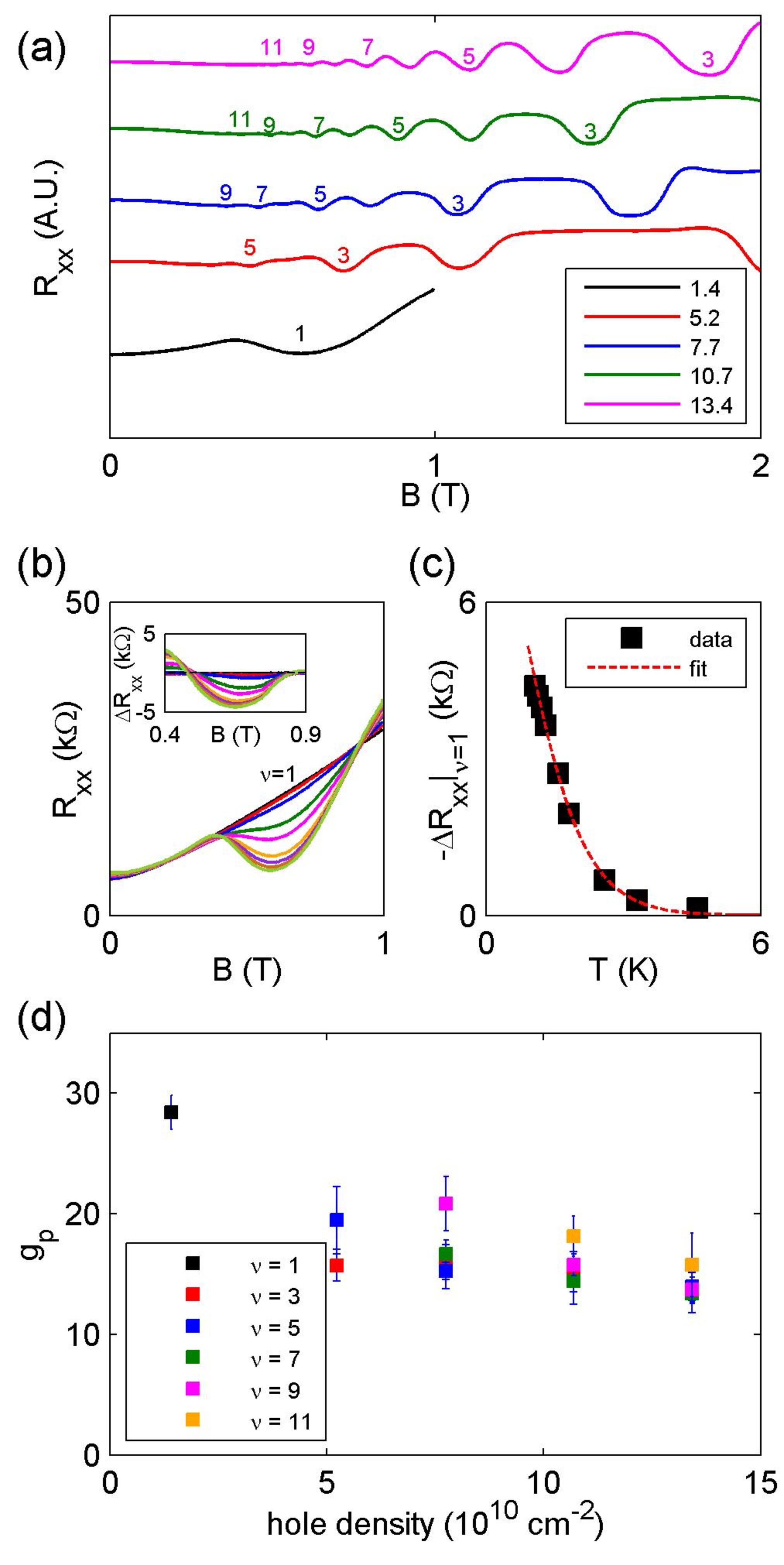}}
\caption{\label{Fig1} (\textbf{a}) $R_{xx}$ at $p=1.4$, $5.2$, $7.7$, $10.7$, and $13.4\times10^{10}$ cm$^{-2}$ at $T=1$ K.  
Minima at odd $\nu$'s are labeled.   The curves are normalized to their zero-field resistance and are offset for clarity.
(\textbf{b})  $R_{xx}$ at $p=1.4\times10^{10}$ cm$^{-2}$ at $T=4.61$, $3.31$,  $2.60$, $1.82$, $1.57$, $1.31$, $1.22$, $1.14$, and $1.08$ K.  
Inset: after subtracting a background obtained by a moving average filter.
(\textbf{c})  Temperature dependence of the SdH oscillation amplitude at $\nu=1$ at $1.4\times10^{10}$ cm$^{-2}$  The red dashed line is a fit to the thermal damping factor.
(\textbf{d})  Extracted $g_p$ as a function of $p$.
}
\end{figure}

Figure 1(a) shows $R_{xx}$ at five densities as a function of magnetic field ($B$) at $T=1$ K.
As $p$ increases, an increasing number of SdH oscillations can be resolved owing to better mobilities at higher densities.
In this density range, $R_{xx}$ shows deeper resistance minima at odd filling factors ($\nu$'s) than at even at high magnetic fields, and only the minima at odd $\nu$'s remain at low magnetic fields and low densities.
For example, at $p=1.4\times10^{10}$ cm$^{-2}$, only the $\nu=1$ minimum is observable.
At $p=7.7\times10^{10}$ cm$^{-2}$, $\nu=6$ at $B\sim0.53$ T is barely resolvable, while $\nu=5$, $7$, and $9$ can still be clearly distinguished.
This behavior is quite different from what is observed in other Ge 2D hole systems, typically with a much higher hole density \cite{Irisawa2003,Morrison2016}.
In fact, we previously observed a crossover from strong minima at even $\nu$'s to strong minima at odd $\nu$'s as $p$ drops below $\sim1.5\times10^{11}$ cm$^{-2}$ \cite{Lu2017}.
This effect can be understood as the Zeeman splitting $g_p\mu_BB$ growing larger than half of the cyclotron gap $\hbar eB/m^*$ below this crossover density.
Here, $\mu_B$ is the Bohr magneton, $\hbar$ is the reduced Planck constant, and $e$ is the elementary charge.
In the density range studied in this work, we conclude that the Zeeman splitting at odd $\nu$'s is the dominant energy scale in a perpendicular magnetic field.
The energy gap at even $\nu$'s is $\hbar eB/m^*-g_p\mu_BB$ and is smaller.
See Ref.~\onlinecite{Lu2017} for a schematic drawing of the Landau level structure in this scenario.

In real 2D systems, each Landau level is broadened by disorder.  
The density of states of a 2D system in a small magnetic field can be calculated by summing over individual disorder-broadened Landau levels and is oscillatory with an energy scale typically associated with the cyclotron gap. 
The SdH oscillations in $R_{xx}$ is directly proportional to this oscillatory density of states \cite{Coleridge1989}.
At finite temperatures, the SdH oscillation amplitude is reduced by a thermal damping factor  $\frac{X}{sinh X}$, where $X$ is $\frac{2\pi^2k_BT}{\Delta}$.  
Here $k_B$ is the Boltzmann constant, and $\Delta$ is the relevant energy gap. 
To extract $\Delta$, one can simply fit the temperature dependence of the SdH oscillation amplitude to the thermal damping factor.
In most cases where Zeeman splitting is small, the SdH analysis is typically applied to obtain the cyclotron gap $\hbar eB/m^*$, which ultimately yields $m^*$.
In the case of interest here, the relevant energy gap is the Zeeman splitting $g_p\mu_BB$.  
The temperature dependence of the SdH oscillation amplitude thus allows us to extract $g_p$.

Figure 1(b) shows the temperature dependence of the SdH oscillation at $\nu=1$.  
As $T$ decreases, the minimum at $\nu=1$ deepens.
To obtained the oscillation amplitude, we pass the data through a moving average filter and remove this slow varying component from the data, as shown in the inset.
We fit the temperature-dependent SdH oscillation amplitude $-\Delta R_{xx}$ to the damping factor to obtain the Zeeman splitting for a given $p$ and $B$, as shown in Fig.~1(c).
The Zeeman splitting is then converted to $g_p$.
For a fixed $p$, there can be multiple $R_{xx}$ minima at odd $\nu$'s, and we deduce $g_p$ for the ones that have not entered the quantum Hall regime.
The density dependence of $g_p$ is shown in Fig.~1(d).
While $g_p$ appears to be slightly $\nu$ dependent, $g_p$ is in general greater than $\sim13$ and increases as $p$ decreases.
At the lowest density $p=1.4\times10^{10}$ cm$^{-2}$, $g_p$ is as large as $\sim28$.
Enhancement of $g_p$ with decreasing density has been observed in other material systems and can be attributed to an interaction effect \cite{Englert1982, Yu2002}.
We previously observed a quantum Hall ferromagnetic transition at $p=2.4\times10^{10}$ cm$^{-2}$ and concluded that at this density the cyclotron gap equals the Zeeman gap \cite{Lu2017}.  
In this case, we obtain $m^*g_p=2$ $m_0$ by evaluating the expressions for the cyclotron gap and the Zeeman splitting.
Based on the results in Fig.~1(d), we can estimate that $g_p\sim20-25$ at the critical density of the quantum Hall ferromagnetic transition.
This leads to $m^*\sim0.08-0.1$  $m_0$, which is consistent with previously measured values of $m^*$ at higher densities.

\begin{figure}[h]
\resizebox{3.3 in}{!}{\includegraphics{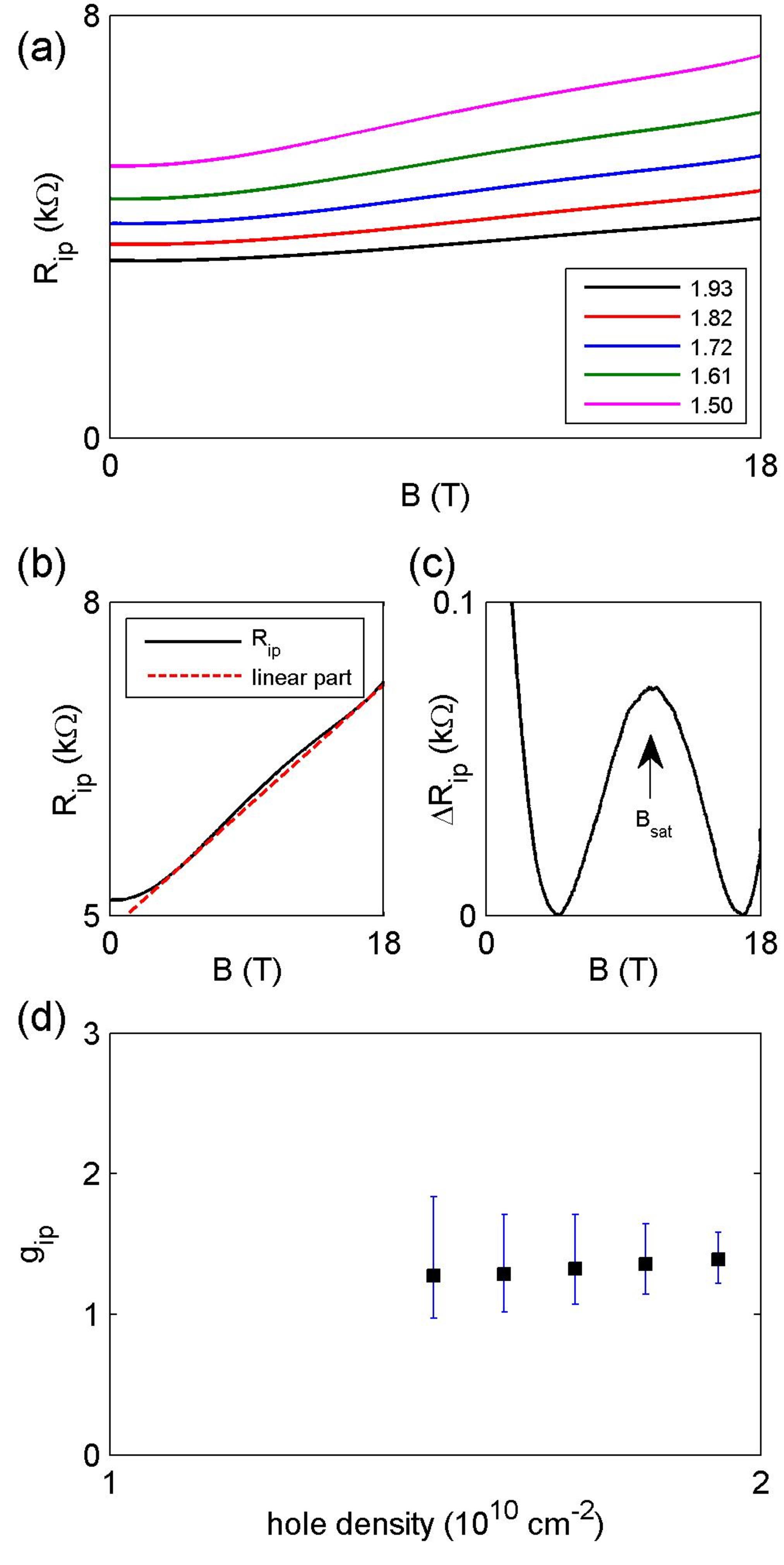}}
\caption{\label{Fig2} (\textbf{a}) $R_{ip}$ at $p=1.50$, $1.61$, $1.72$, $1.82$, and $1.93\times10^{10}$ cm$^{-2}$ at $T=0.3$ K.
(\textbf{b}) $R_{ip}$ at $p=1.50\times10^{10}$ cm$^{-2}$ with the linear component shown in red dashed line.
(\textbf{c}) After removing the linear component, $\Delta R_{ip}$ has a peak at $B\sim11$ T, indicating full spin saturation at this magnetic field.
(\textbf{d}) Extracted $g_{ip}$ as a function of $p$.
}
\end{figure}

For a Ge 2D hole system, the effective $g$ factor is expected to be highly anisotropic due to the symmetry of the valance band as well as quantum confinement and strain, which set a preferential orientation for hole spins \cite{Martin1990,Lin1991,Winkler2000, Winkler2003}.
However, the in-plane component $g_{ip}$ of Ge 2D holes is difficult to measure quantitatively.
The main experimental tool for measuring $g_{ip}$ is to search for a jut in the $R_{ip}$ data, which indicates full spin polarization in the presence of an in-plane magnetic field \cite{Okamoto1999}.
At the characteristic magnetic field, $B_{sat}$, where the resistance jut occurs, the Zeeman splitting $g_{ip}\mu_BB$ equals the Fermi energy of the now spinless 2D system, $p/(m^*/2\pi\hbar^2)$.
Together with the known density, $B_{sat}$ can be converted to $g_{ip}$.
The primary challenge in measuring $g_{ip}$ of Ge 2D holes arises from the relatively small magnitude of $g_{ip}$ and the typically large values of hole density.
A substantially high magnetic field is required for such experiments and is often out of reach.
Our HFET device permitted access to a desired low-density regime $(10^{10}$ cm$^{-2})$ electrically.
By measuring $R_{ip}(B)$ in an $18$-T magnet in the National High Magnetic Field Laboratory, we were able to observe protrusions in resistance data and extract $g_{ip}$.

In Fig.~2(a), we show $R_{ip}$ at five densities as a function of in-plane magnetic field at $T=0.3$ K.  
A weak jut in the resistance can be seen in the lowest-density curve, while the jut weakens as $p$ increases.
The jut moves toward higher $B$ as $p$ increases, consistent with the expected density dependence.
The fact that $R_{ip}$ does not saturate beyond $B_{sat}$ is well understood to be a finite thickness effect \cite{DasSarma2000}.
We subtract a linear component from $R_{ip}(B)$, as shown in Fig.~2(b) for the lowest density curve, to make the jut apparent and thus creating a peak in resistance, as shown in Fig.~2(c).
The location of the resistance peak is defined as $B_{sat}$ and is used to extract $g_{ip}$.
We take $m^*\sim0.09$~$m_0$ in the evaluation of $g_{ip}$, based on the our estimation in this density range discussed above.
The extracted $g_{ip}$ is shown in Fig.~2(d) as a function of $p$.
We note that the density range is limited, as we did not observe clear resistance peaks at $p>2\times10^{10}$ cm$^{-2}$.
Nevertheless, the small magnitudes of $g_{ip}$ observed in this density range confirms and quantifies the expected anisotropy in the effective $g$ factor.
Taking $g_{ip}\sim1.3$, we obtain an anisotropy $g_p/g_{ip}$ of $\sim22$ at $p\sim1.5\times10^{10}$ cm$^{-2}$.

In summary, we present our measurements of $g_p$ and $g_{ip}$ of Ge 2D holes in the low-density regime.
A very strong anisotropy is observed, with $g_{ip}\sim1.3$ and $g_p$ at least an order of magnitude higher.
As $p$ decreases, $g_p$ shows interaction-induced enhancement, and reaches a value as large as $28$.
These measured values are important material parameters that can be used for estimating device performance and understanding the spin physics of Ge 2D holes.  

We thank A. Suslov, H. Baek, G. Jones, and T. Murphy for their assistance in experiments.  This work has been supported by the Division of Materials Sciences and Engineering, Office of Basic Energy Sciences, U.S. Department of Energy (DOE).  This work was performed, in part, at the Center for Integrated Nanotechnologies, a U.S. DOE, Office of Basic Energy Sciences, user facility.  Sandia National Laboratories is a multi-mission laboratory managed and operated by National Technology and Engineering Solutions of Sandia, LLC., a wholly owned subsidiary of Honeywell International, Inc., for the U.S. Department of Energy's National Nuclear Security Administration under contract DE-NA-0003525.  A portion of this work was performed at the National High Magnetic Field Laboratory, which is supported by National Science Foundation Cooperative Agreement No. DMR-1157490 and the State of Florida.  The work at NTU was supported by the Ministry of Science and Technology (103-2112-M-002-002-MY3 and 105-2622-8-002-001).

\end{document}